\headline={\ifnum\pageno=1\firstheadline\else
\ifodd\pageno\rightheadline \else\leftheadline\fi\fi}
\def\firstheadline{\hfil}
\def\rightheadline{\hfil}
\def\leftheadline{\hfil}
        \footline={\ifnum\pageno=1\firstfootline\else\otherfootline\fi}
\def\firstfootline{\rm\hss\folio\hss}
\def\otherfootline{\hfil}

\font\tenrm=cmr10

\font\elevenbf=cmbx10 scaled\magstep 1
\font\elevenrm=cmr10 scaled\magstep 1
\font\elevenit=cmti10 scaled\magstep 1

\font\ninerm=cmr9

\nopagenumbers
\hsize=6.0truein
\vsize=8.5truein
\parindent=1.5pc
\baselineskip=10pt
\centerline{\elevenbf SUSY AND THE DECAY $H^0_2\rightarrow gg$ 
\footnote*{{\ninerm\baselineskip=11pt 
Talk presented at the MRST-96, University of Toronto,
Toronto, Canada, May 9-10, 1996, UQAM-PHE-96/03.}}}
\vglue.5cm 
\centerline{\elevenrm HEINZ K\"ONIG}
\baselineskip=12pt
\centerline{\elevenit D\'epartement de Physique}
\baselineskip=12pt
\centerline{\elevenit Universit\'e du Qu\'ebec \`a Montr\'eal}
\baselineskip=12pt
\centerline{\elevenit C.P. 8888, Succ. Centre-Ville, Montr\'eal}
\baselineskip=12pt
\centerline{\elevenit Qu\'ebec, Canada H3C 3P8}
\vglue 0.5cm
\centerline{\tenrm ABSTRACT}
\vglue 0.5cm
  {\rightskip=3pc
 \leftskip=3pc
 \tenrm\baselineskip=12pt
 \noindent
In this talk I present a detailed SUSY QCD calculation 
of the decay rate of the lightest
Higgs boson $H^0_2$\ into two gluons, where all quarks
and scalar quarks are taken within the relevant
loop diagrams. I include the mixing of all
the three generations of the scalar
partners of the left and right handed quarks
and show that their contribution is 
comparable to the quark contribution in the MSSM 
for small values of the soft SUSY 
breaking parameter $m_S$. Furthermore 
in the MSSM the contribution from the bottom 
quark becomes as large as the top quark contribution
for large $\tan\beta$\ and large Higgs masses.
As a result, the two gluon decay rate of $H_2^0$\ is much
larger than the two gluon decay rate of an equal mass
standard model Higgs boson.
I further compare the decay mode 
of $H^0_2\rightarrow gg$\
to the similar decay modes of $H^0_2\rightarrow c\overline c$\ 
including one loop QCD corrections
and show that in some cases 
$\Gamma(H^0_2\rightarrow gg)$\ is even higher than 
$\Gamma(H^0_2\rightarrow c\overline c)$.
\vglue 0.5cm}\noindent 
{\elevenbf 1. Introduction}
\vglue 0.5cm
\baselineskip=12pt
\elevenrm
The Higgs boson is the last particle in the standard
model (SM), which yet lacks any experimental evidence.
Its discovery therefore is of great importance. 
The instruments of discovery will be LEP
if the Higgs mass is smaller than the Z boson mass and LHC
for higher masses.
While for a Higgs mass smaller than twice of the gauge
boson mass the most important decay modes for its
discovery will be $H\rightarrow q\overline q$\
(here $q=c,b$) and $H\rightarrow \gamma\gamma$\
and to some extent $H\rightarrow gg$,
it will be the decay into two W or Z bosons
for higher masses of the Higgs boson.
\hfill\break\indent
It is well known that the SM is not a sufficiant model
when considering unification theories. The favourite
model beyond the SM is its minimal supersymmetric extension
(MSSM) [1]. The content of Higgs particles in the MSSM
is quite different than the one of the SM: it contains
two scalar Higgs bosons $H_1^0,H^0_2$, 
one pseudo-scalar $H_3^0$\ and one
charged scalar $H^\pm$. The most important point is that
the mass of the lightest Higgs particle 
$m_{H^0_2}$\ has to be
smaller than the Z boson mass at tree level and is
enhanced to a maximum value of around 130 GeV when
loop corrections are included [2], thus making
a SUSY Higgs boson more experimentally reachable. 
\hfill\break\indent
For a Higgs boson far less massive than the Z
boson, the $H^0_2\rightarrow \gamma\gamma$\ is
the  most important decay mode and was analyzed 
in [3] (for the SM) and in [4] (for the MSSM).
For values of the Higgs mass up to twice 
the W and Z boson masses the decays 
$H^0_2\rightarrow q\overline q$\ and $H^0_2\rightarrow gg$\
will become more important.
The QCD corrections to the first decay mode within 
the SM were considered in [5] (and references therein)
and within the MSSM in [6]. The second decay mode
was considered in [7] and two loop QCD corrections within the
SM were considered in [8] and found to be relatively
large: about 60\%. In this talk I show that the 
MSSM leads to a much higher Higgs into two gluons 
decay rate than the SM for some supersymmetric parameters,
making this decay mode more interesting.
\hfill\break\indent
It will be difficult to measure $\Gamma(H^0_2\rightarrow gg)$\
due to QCD jet background, although it might be experimental
measurable at future $e^+e^-$-- colliders [9]. 
Therefore it is important to consider
all kind of models in regards to this decay mode and as I will
show the scalar quarks contribution can be several 
tens of per cent compared to the quark contribution
and also to the $H^0_2\rightarrow
c\overline c$\ decay rate 
for some SUSY parameter space
after summing over all scalar quarks.
\hfill\break\indent
In the next section I present the results 
and discuss them in the third section.
In the calculation I include the mixing of all
scalar partners of the left and right handed
quarks, which is expected to be of importance
in the top quark sector due to the high top quark mass
of $180$\ GeV reported by the CDF and $D\emptyset$\ groups [10]. In the 
bottom quark sector I also include one loop effects.
As a surprise I also find, that the mixing is 
not negligible in the charm and strange quark sector independant
of the value for $\tan\beta$, the ratio of the Higgs vacuum
expectation values (vev's). In this talk I only present the
results and refer the interested reader for detailed calculations
to [11].  
\vglue 0.5cm\noindent
{\elevenbf 2. SUSY QCD Corrections to $H^0_2\rightarrow gg$}
\vglue 0.5cm
In the MSSM there are strong relations among the
masses and mixing angles of the Higgs bosons. Given two
values e.g. $\tan\beta=v_2/v_1$\ (the ratio of the vacuum 
expectation values)
and the light Higgs boson
mass $m_{H^0_2}$\ all the other masses and angles
including one loop corrections can be obtained analytically 
as presented in [11].
\hfill\break\indent  
In the SM the decay mode of the Higgs boson into two gluons
occur via one loop diagrams with all quarks taken within the
loop as shown in Fig.1.
The final amplitude is finite and the result is given by:
$$\eqalignno{iM_q=&
+g_2{{g^2_s}\over{(4\pi)^2}}{{m_Z^2}\over
{m_W}}{4\over{q^2}}(p_1p_2g_{\alpha\beta}-
p_{1\beta}p_{2\alpha})\epsilon^{\ast\alpha}_{p_1}
\epsilon^{\ast\beta}_{p_2}T_q&(1)\cr
T_q=&\sum\limits_q{{m_q^2}\over{m_Z^2}}K^{qH_2^0}
\lbrack 2-(1-4\lambda_q)I_q\rbrack\cr}$$
with $q^2=m_{H_2^0}^2$\ on mass shell and 
$K^{uH_2^0}=\cos\alpha/\sin\beta$\ and $K^{dH_2^0}=
-\sin\alpha/\cos\beta$\ and the function $I_q$\ defined by
$$\eqalignno{
I_q=&\biggl\lbrace\matrix{
-2\lbrack \arcsin({1\over{2\sqrt{\lambda_q}}})
\rbrack^2,& 1/4\leq\lambda_q\cr\lbrack \ln({{r_+}\over{r_-}})^2
-\pi^2\rbrack/2+i\pi\ln({{r_+}\over{r_-}}),&\lambda_q < 1/4
\cr}&(2)\cr
r_\pm=&1\pm (1-4\lambda_q)^{1\over 2}\cr}$$
with $\lambda_q=(m_q/m_{H_2^0})^2$. 
Before I present the results of the scalar quarks
contribution to the Higgs decay into two gluons
I first want to comment on their mass matrices in
the MSSM. The mixing term of the scalar partners
of the left and right handed quarks is proportional
to the quark masses and 
hence was neglected 
before the top quark was discovered as very heavy.
In the calculation here I include the
mixing of all scalar quarks and present the 
result in their mass eigenstates,  
that is instead of the current eigenstates
$\tilde q_{L,R}$\ I 
work with the mass eigenstates
$$\tilde q_1=cos\Theta_q\tilde q_L+\sin\Theta_q\tilde q_R\qquad
\tilde q_2=-\sin\Theta_q\tilde q_L+\cos\Theta_q\tilde q_R\eqno(3)$$
Here $q$\ stands for all three generations of the scalar up and
scalar down quarks. A detailed description of their mass matrices
including one loop corrections for the scalar down quarks is
presented in [11].
In the calculation it turns out that the mixing of the
first generation is negligible as expected, whereas
in the second generation $sin\Theta_q\simeq 0.1-0.5$\
(the last value only for $\tan\beta\gg 1$)
and therefore not negligible. In the third generation
$\sin\Theta_q\simeq 1/\sqrt{2}$\ due to the heavy top
quark mass. For the scalar bottom quark the mixing angle
only becomes that big when $\tan\beta\gg 1$. 
\hfill\break\indent
\vskip6cm
$$\vbox{\settabs2\columns\tenrm
\+ Fig.1: The penguin diagram with up and down   
&\quad
Fig.2: The penguin diagrams with scalar up\cr
\+ \phantom{Fig.1: }quarks within the loop   
&\quad \phantom{Fig.2: }and down quarks within the loop\cr
}$$\indent
In the MSSM we have to add up the two diagrams shown
in Fig.2. After summation the amplitude is finite
and the result is given by:
$$\eqalignno{iM_{\tilde q}=&-{{g_2}\over{\cos\Theta_W}}
{{g_s^2}\over{(4\pi)^2}}{{4m_Z}\over{q^2}}
(p_1p_2g_{\alpha\beta}-p_{1\beta}p_{2\alpha})
\epsilon^{\ast\alpha}_{p_1}\epsilon^{\ast\beta}_{p_2} 
T_{\tilde q}&(4)\cr
T_{\tilde q}=&\sum\limits_q\bigl\lbrace
\lbrack\cos^2\Theta_q K^{\tilde q H_2^0}_{11}
+\sin^2\Theta_q K^{\tilde q H_2^0}_{22}
+2\sin\Theta_q
\cos\Theta_q K^{\tilde q H_2^0}_{12}
\rbrack(1+2\lambda_{\tilde q_1}
I_{\tilde q_1})\cr
&+\lbrack\sin^2\Theta_q K^{\tilde q H_2^0}_{11}+\cos^2\Theta_q
K^{\tilde q H_2^0}_{22}
-2\sin\Theta_q\cos\Theta_q 
K^{\tilde q H_2^0}_{12}
\rbrack(1+2\lambda_{\tilde q_2}I_{\tilde q_2})\bigr\rbrace\cr
K^{\tilde u H_2^0}_{11}=&-({1\over 2}-e_u s_W^2)\sin(\alpha
+\beta)+\left ({{m_u}\over{m_Z}}\right )^2{{\cos\alpha}
\over{\sin\beta}}\cr
K^{\tilde u H_2^0}_{22}=&-e_u s_W^2\sin(\alpha+\beta)+
\left ({{m_u}\over{m_Z}}\right )^2{{\cos\alpha}
\over{\sin\beta}}\cr
K^{\tilde u H_2^0}_{12}=&K^{\tilde u H_2^0}_{21}=
{1\over 2}{{m_u}\over{m_Z^2}}{1\over{\sin\beta}}(A_u\cos\alpha
+\mu\sin\alpha)\cr
K^{\tilde d H_2^0}_{11}=&({1\over 2}+e_d s_W^2)\sin(\alpha
+\beta)-\left ({{m_d}\over{m_Z}}\right )^2
{{\sin\alpha}\over{\cos\beta}}\cr
K^{\tilde d H_2^0}_{22}=&-e_d s_W^2\sin(\alpha+\beta)
-\left ({{m_d}\over{m_Z}}\right )^2
{{\sin\alpha}\over{\cos\beta}}\cr
K^{\tilde d H_2^0}_{12}=&K^{\tilde d H_2^0}_{21}=
-{1\over 2}{{m_d}\over{m_Z^2}}
{1\over{\cos\beta}}(A_d\sin\alpha+\mu\cos\alpha)
\cr}$$
with $s_W^2=\sin^2\Theta_W$\ and again $q^2=m_{H_2^0}^2$\
on mass shell. Note that the non diagonal
terms $K_{12}^{\tilde q H_2^0}$\ in $T_{\tilde q}$\  
only contribute when
the scalar mass eigenstates differ, which mainly is the
case for the third generation. Note also that $T_{\tilde q}$\
is identical to $0$\ if all scalar quarks have equal masses. 
The amplitudes in eq.(1) and eq.(4) lead to the 
following decay rate:
$$\Gamma(H_2^0\rightarrow gg)={{\alpha\alpha_s^2}\over
{8\pi^2\sin^2\Theta_W\cos^2\Theta_W}}
{{m_Z^2}\over{m_{H_2^0}}}
\vert T_q-T_{\tilde q}
\vert^2\eqno(5)$$
If $T_{\tilde q}$\ is set to $0$\
eq.(5) reproduce eq.(2.29) given in [12]. 
\hfill\break\indent
In the next section I will discuss the results of the
lightest supersymmetric Higgs boson into two gluons decay rate
obtained in eq.(8). 
\vglue 0.5cm\noindent
{\elevenbf 3. Discussions}
\vglue 0.5cm
To see how big the contribution of the scalar quarks
compared to the quarks is 
I plot in Fig.3 the ratio  
$\Gamma^{\tilde q+q}/\Gamma^q$\ of the decay
rate $\Gamma(H_2^0\rightarrow gg)$\ as function of
the soft SUSY breaking scalar mass $m_S$\ for a 
fixed value of $\mu=250$\ GeV, the bilinear Higgs mass term, and 
two different values of the Higgs mass $m_{H_2^0}=60$\ GeV
and $120$\ GeV
and three different values of $\tan\beta=3$\ (solid
line), $10$\ (dashed line) and $60$\ (dotted line).
Higher values of $\tan\beta$\ are preferred in
superstring inspired $E_6$\ and $SO(10)$\ models.
For $\Gamma^q$\ and  
$\Gamma^{\tilde q+q}$\ I have taken
$T_q$\ as given in eq.(2) with the couplings $K^{qH_2^0}$,
that is including the large enhancements (relative to the 
SM) due to large $\tan\beta$.
As a result I have that for small values of $m_S$\ 
the scalar quarks contribute even more than the quarks,
although their contribution decrease rapidly and 
remains only a few per cent for $m_S>600$\ GeV.
For $\tan\beta=60$\ the scalar quarks contribution 
diminishes the ratio for $m_S<350$\ GeV and enhances it
for higher values.  
The influence of $\mu$\ is very small for small 
$\tan\beta$\ and becomes more important for
very high $\tan\beta$\ values. For small values
of $\tan\beta$\ higher values of $\mu$\ 
enhance the decay rate a little bit. For high values
of $\tan\beta$\ it is the other way around and the
differences are larger.
A negative value 
for $\mu$\ leads to a bit smaller values 
of the decay rate. 
\hfill\break\indent
\phantom{Scheisse}
\vskip6cm
$$\vbox{\settabs2\columns\tenrm
\+ Fig.3: The ratio of $\Gamma^{\tilde q+q}/\Gamma^q$\ as explained in  
text.&\quad
Fig.4: The ratios of $\Gamma^{\tilde q+q}/\Gamma^q$\ and
$\Gamma^{\tilde q+q}/\Gamma^{c\overline c}$\ for \cr
\+ \phantom{Fig.3: }Except for $\tan\beta=3$\ the upper curves are  
&\quad \phantom{Fig.4: }$m_{H_2^0}=60$\ GeV as explained in text.\cr
\+ \phantom{Fig.3: }for $m_{H_2^0}=60$\ GeV.\quad \cr}$$\indent
In Fig.4 I have plotted the ratio of the Higgs into two
gluons decay rate of the MSSM compared to the SM, that
is $\Gamma^{\tilde q+q}/\Gamma^q$, where I have taken
$\Gamma^q$\ as it is in the SM that is without 
the couplings $K^{qH_2^0}$, whereas I included them
in $\Gamma^{\tilde q+q}$.
For the Higgs mass I have taken $60$\ GeV. 
As a result I have that
for scalar masses smaller than $500$\ GeV the Higgs
into two gluons decay rate is enhanced by several
tens of per cents in the MSSM and gives the same
result than the SM for higher values of the scalar
masses. As in Fig.3 for $\tan\beta=60$\ and $m_S<350$\
GeV the decay is diminished. 
\hfill\break\indent
In Fig.5 I have done the same as in Fig.4 but for
a Higgs mass of
$m_{H_2^0}=120$\ GeV. Here the results are quite different
than in Fig.4. For $\tan\beta=3$\ and $m_S\ge 600$\ GeV the 
pseudo Higgs obtains a negative mass squared. For $\tan\beta=10$\
the same happens for a small region when $m_S\approx 650$\ GeV. 
As a result I have that in the MSSM the Higgs into
two gluon decay rate is enhanced by several tens of
per cent for $\tan\beta=3$, by a factor of $2-3$\ 
for $\tan\beta=10$\ and $m_S<300$\ GeV and 
even by an order of magnitude for $\tan\beta=60$\
with the highest contribution for a scalar mass around
$550$\ GeV.
\hfill\break\indent
As I have shown in Fig.3 the scalar quarks 
decouples for $m_S>600$\ GeV. The reason why the
branching ratio as shown in Fig.5 is still larger
than $1$\ even for higher values of the scalar mass
is that $\Gamma^q$\ is quite different in the
MSSM with the couplings $K^{qH_2^0}$\ than it
is in the SM without these couplings.
In the SM the main contribution is from
the heavy top quark and a few per cent from the
bottom quark. The contribution of the other quarks
are negligible due to their small masses. 
In the MSSM the bottom quark contribution becomes
as important as the top quark contribution for
large $\tan\beta$\ values eg. the ratio $\Gamma^q_{\rm SM}/
\Gamma^q_{\rm MSSM}$\ becomes very small 
depending on the size and sign of $\sin\alpha$, 
which becomes relative large around 
$\pm 0.5$\ and thus leading
to large values of $\Gamma^{\tilde q+q}/\Gamma^q$\ as
seen in Fig.5. For a small Higgs mass of $60$\ GeV
as I have taken in Fig.4 $\sin\alpha$\ remains always
smaller than around $-2\times 10^{-2}$\ and therefore
keeps the bottom quark mass contribution as small as
in the SM.
\hfill\break\indent 
Some curves in Fig.4--5 start at different values of
$m_S$\ because, for
values of $m_S$\ higher than $600$\ GeV 
I obtain an unphysical negative mass squared
for the pseudo particle $H_3^0$\ if $\tan\beta=3$;
for $\tan\beta=10$\ the unphysical region is when
$m_S\simeq 650$\ GeV; whereas for $\tan\beta=60$\
$m_{H_3^0}$\ is physical for all $m_S$. 
\hfill\break\indent 
\vskip6cm
$$\vbox{\settabs2\columns\tenrm
\+ Fig.5: The same as in Fig.4 with $m_{H_2^0}=120$\   
&\quad
Fig.6: The same as Fig.4 but as a function of\cr
\+ \phantom{Fig.5: }GeV. The upper curves are for 
$\Gamma^{\tilde q+q}/\Gamma^{c\overline c}$. 
&\quad \phantom{Fig.6: }$m_{H_2^0}$. 
The upper curves at the higher  
\cr
\+&\quad \phantom{Fig.6: }Higgs masses. 
are for $\Gamma^{\tilde q+q}/\Gamma^{c\overline c}$.\cr
}$$\indent
A negative eigenvalue
of the scalar bottom quark mass also occurs if $m_S<200$\
GeV  for $\tan\beta=3$\ and $10$\
or $m_S<300$\ GeV for $\tan\beta=60$.
Here the parameter $c$, which enters in the one loop corrections
to the scalar down quark masses,
is of importance, neglecting it would allow 
us to use $m_S$\ as small as 100 GeV (for $\tan\beta=3$\ ) 
without running into one negative mass eigenvalue of
the scalar bottom quark mass, with the result that
$\Gamma^{\tilde q+q}$\ can become much larger than $\Gamma^q$.
Unfortunately even for smaller values for $c\simeq -0.5$\
I obtain negative values with such a small scalar mass.
Since $c$\ cannot be neglected when including loop corrections 
I excluded those regions in the figures.
\hfill\break\indent
In Fig.4 and Fig.5 I also have plotted the ratio of
the decay rates $\Gamma(H_2^0\rightarrow gg)/\Gamma(H_2^0\rightarrow 
c\overline c)$.
 For the decay rate $\Gamma(H_2^0
\rightarrow b\overline b, c\overline c)$\ I used 
the tree result including the SM QCD corrections as 
given in eq.(8) of Ref. [5]
with the changes of the tree level couplings within the MSSM.
I did not include the SUSY QCD
correction, because they are far
smaller than the SM QCD correction
as I have shown in [6]. There I showed 
that for $\tan\beta=1$\ SUSY QCD corrections 
do not contribute at
all to this decay mode ($\sin(\alpha+\beta)=0$)\ and presented
the results in the limit of $\tan\beta\gg 1$.
There I did not include the mixing of the scalar charm
and bottom quarks, but since $m_{\tilde c_1}\approx m_{\tilde c_2}$\
even a large mixing angle will not change the results for
the decay rate $\Gamma(H_2^0\rightarrow c\overline c)$
presented there.
This might not be true for $\Gamma(H_2^0\rightarrow
b\overline b)$\ which I did not consider here 
since it is much higher than $\Gamma(H_2^0\rightarrow gg)$, 
by a factor of at least 50. Therefore
in Fig.4 and Fig.5 I only compared the Higgs into two gluons
decay rate with the one to the charm- anti-charm quarks.
In Fig.4, the dependence of the ratio to
the scalar mass $m_S$\ is very small, since $\Gamma^{\tilde q+q}$\
becomes very small and the dependence of $\Gamma^{c\overline c}$\
to $m_S$\ is only via the angles $\cos^2\alpha/\sin^2\beta$,
which is compensated by the $K^{tH_2^0}$\ coupling in the $T_q$\
term. For a very large scalar mass the ratio remains constant
with a value of around $0.31$\ independant of $\tan\beta$.
A quite different result I obtain in Fig.5, especially again for
$\tan\beta=60$, for the same reason as explained above. The
shape of the figures is quite similar compared to the
ratio $\Gamma^{\tilde q+q}/\Gamma^q$. For scalar masses much
higher than $1$\ TeV the ratio remains constant with a value
of around $1$\ independant of $\tan\beta$.
\hfill\break\indent
Finally in Fig.6 I show the influence of the
Higgs mass to the decay rate $\Gamma(H_2^0\rightarrow gg)$\
for a fixed value of $\mu=250$\ GeV and $m_S=300$\ GeV
and three different values of $\tan\beta=3,10$\ and $60$.
In the case $\tan\beta=3$\ I obtain negative values for
the mass squared of the pseudo Higgs $H_3^0$\ in the range
of $95<m_{H_2^0}<105$\ GeV which therefore has to be
excluded.  As a result I have that $\Gamma^{\tilde q+q}/
\Gamma^q$\ is weakly dependant of the Higgs mass,
not so $\Gamma^{\tilde q+q}/\Gamma^{c\overline c}$, which
shape is basically dominated by 
$\cos\alpha$\ and $\sin\beta$.
\vglue 0.5cm\noindent
{\elevenbf 4. Conclusion}
\vglue 0.5cm
In this talk I presented the corrections to the
lightest MSSM Higgs boson decay into two gluons
when scalar quarks are taken within the loop.
I included in my calculation the mixing of all
scalar quarks although it only becomes important
for the second and third generation. I
have shown that scalar quarks lead to a 
decay rate of the same order
as the quarks in the SM
for values of $m_S$\ smaller than $600$\ GeV. 
In the SM the largest
contribution comes from the top quark due to the 
$m_q^2$\ in $T_q$. In the MSSM the $T_{\tilde q}$\ are
of the same order for all scalar quarks and therefore
contribute many more terms to $\Gamma(H_2^0\rightarrow gg)$\
than the SM alone. Furthermore in the MSSM the $T_q$\
can become much larger than in the SM for $\tan\beta\gg 1$\
and large negative or positive $\sin\alpha$.
I also have shown that the Higgs into two gluon
decay rate can become even larger than the decay into
charm- anti-charm quarks for $\tan\beta=3$\ and
the Higgs mass larger
than around $80$\ GeV and for $\tan\beta=10$\ and $60$\ 
and the Higgs mass larger than the Z boson mass,
but still remains more
than a factor of $50$\ smaller than its decay into bottom-
anti-bottom quarks.
\hfill\break\indent
Although the decay of the Higgs into two gluons will
be difficult to measure it is of importance to know
how big the influence of models beyond the SM might be.
Furthermore it might be measurable at future $e^+e^-$--
colliders [9].
The amplitudes given here can also be used when considering
Higgs production in hadron colliders
via gluon fusion with following decay in heavy leptons [13].
\vglue 0.5cm\noindent
{\elevenbf References}
\vglue 0.5cm
\item{[\ 1]}H.E. Haber and G.L. Kane, Phys.Rep.{\elevenbf 117}(1985)75.
\item{[\ 2]}H. Haber and R. Hempfling, Phys.Rev.Lett.{\elevenbf 66}
(1991)1815; Y. Okada et al., Prog.Theor.Phys.{\elevenbf 85}(1991)1;
Phys.Lett.{\elevenbf 262B}(1991)54; J. Ellis et al., ibid {\elevenbf 257B}
(1991)83; for a more general analysis leading to a higher mass
limit of about 150 GeV see G.L. Kane et al., Phys.Rev.Lett.
{\elevenbf 70}(1993)2686.
\item{[\ 3]} L. Resnick, M.K. Sundaresan and P.J. Watson,
Phys.Rev.{\elevenbf D8}(1973)172; A.J. Vainstein et al.,
Yad.Fiz.{\elevenbf 30}(1979)1368
[Sov.J.Nucl.Phys.{\elevenbf 30}(1979)711].
\item{[\ 4]}R. Bates, J.N. Ng and P. Kalyniak, Phys.Rev.
{\elevenbf D34}(1986)172.
\item{[\ 5]}P. Kalyniak et al., Phys. Rev.{\elevenbf D43}(1991)3664.
\item{[\ 6]}H. K\"onig, Phys.Rev.{\elevenbf D47}(1993)4995.
\item{[\ 7]}F. Wilczek, Phys.Rev.Lett.{\elevenbf 39}(1977)1304;
H.M. Georgi et al., Phys.Rev.Lett.{\elevenbf 40}
\hfill\break (1978)692;
J. Ellis et al., Phys.Lett.{\elevenbf 83B}(1979)339; T. Rizzo,
Phys.Rev.{\elevenbf D22}\hfill\break (1980)178.
\item{[\ 8]}A. Djouadi, M. Spira and P.M. Zerwas, Phys.Lett.
{\elevenbf 264B}(1991)440; M. Spira et al, Nucl.Phys.{\elevenbf B453}(1995)
17.
\item{[\ 9]}Proceedings of the Workshop "$e^+e^-$\ Collisions
at 500 GeV: The Physics Potential", DESY Report 92-123A; August
1992, P. Zerwas, ed.
\item{[10]} F. Abe et al., Phys.Rev.Lett{\elevenbf 74}(1995)2626; 
S. Abachi et al., ibid, 2632. 
\item{[11]}H. K\"onig, Z.Phys.{\elevenbf C69}(1996)493.
\item{[12]}J.F. Gunion and et al., {\elevenit The Higgs Hunter's
Guide}, (Addison-Wesley, Redwood City, CA, 1990).
\item{[13]}\underbar{M. Boyce}, M.A. Doncheski and H. K\"onig, 
these proceedings.
\vfill\break\noindent
\eject\bye